\documentclass[pdflatex,sn-mathphys-num,iicol]{sn-jnl}

\usepackage{graphicx}%
\usepackage{multirow}%
\usepackage{amsmath,amssymb,amsfonts}%
\usepackage{amsthm}%
\usepackage{mathrsfs}%
\usepackage[title]{appendix}%
\usepackage{xcolor}%
\usepackage{textcomp}%
\usepackage{manyfoot}%
\usepackage{booktabs}%
\usepackage{algorithm}%
\usepackage{algorithmicx}%
\usepackage{algpseudocode}%
\usepackage{listings}%
\usepackage{siunitx}%
\usepackage{hyperref}

\theoremstyle{thmstyleone}%
\theoremstyle{thmstyletwo}%

\theoremstyle{thmstylethree}%

\begin{document}

\title[Pyroelectricity: A Brief History of its Discovery and
Physical
Principles --
an Overview]{Pyroelectricity: A Brief History of its Discovery and
Physical
Principles --
an Overview}


\author*[1]{\fnm{Rüdiger
G.}\sur{Ballas}}\email{ruediger.ballas@wb-fernstudium.de}

\author[1]{\fnm{Nataliya} \sur{Koev}}\email{nataliya.koev@wb-fernstudium.de}


\affil*[1]{\orgdiv{Department of Engineering Sciences}, \orgname{Wilhelm
Büchner Hochschule},
\orgaddress{\street{Hilpertstrass 31}, \city{Darmstadt}, \postcode{64295},
\state{Hesse},
\country{Germany}}}

%


\abstract{This paper deals with the historical development and physical
mechanisms of
pyroelectricity, a phenomenon whose roots date back over 2000 years to ancient
times. This paper is aimed at students and engineers as
a concise
introduction
to
the subject area. While the attractive effect of heated tourmaline was already
described
by Theophrastus, scientific systematization did not occur until the 18th
century by researchers, such as Aepinus and Canton, who identified the effect
of
electrical polarization resulting from temperature changes. This essay
highlights the path from early analogies to magnetism to the modern
crystallographic description by Haüy and Thomson. In the physics section,
pyroelectricity is defined as the temperature dependence of the spontaneous
polarization 
in anisotropic solids.
At the microscopic level, the permanent dipole moment
of an elementary cell is described by the vector
sum of individual moments.
It is mathematically demonstrated that the macroscopic spontaneous polarization
correlates with the surface charge density and
is
linked to the temperature change 
via the pyroelectric
coefficient. 
A distinction was made between the primary pyroelectric effect and the
secondary
effect resulting from the thermal deformation of the crystal. Finally, the
renaissance of this field of research through the development of modern
infrared detectors and ferroelectric materials in the 20th century was
highlighted.}




\maketitle

\section{Introduction}\label{sec1}

Pyroelectricity is one of the earliest known electrical
phenomena and is 
an important basis for infrared sensors and thermoelectric
applications. Despite its technical relevance, specialist literature
rarely provides a concise, historically contextualized account of
physical principles. This paper is aimed at
physics and engineering students, as well as specialists in related
disciplines, and offers a structured overview of the subject area. It
pursues a two-fold objective: in the historical section
(Section~\ref{sec:Entdeckung})
the
development from antiquity to the 20th century is traced, whereas in the
physical section (Section~\ref{sec:Physik}) the
microscopic and macroscopic
fundamentals are systematically derived and linked to material parameters. This
work is intended as a supplementary introduction to the
standard works of Lang~\cite{Lang2005} and Katzir~\cite{Katzir2006}.

\section{The discovery of pyroelectricity}
\label{sec:Entdeckung}

When piezoelectricity was discovered in 1880, pyroelectricity was
known and researched for more than a century. Unlike piezoelectricity, which
was detected in targeted laboratory experiments by observing the electrical
effect under mechanical pressure, the discovery of pyroelectricity arose from
chance observations of heated tourmaline.

The attractive effect of heated tourmaline was known even before scientific
research and dates back to more than 2000 years. It was probably the Greek
philosopher Theophrastus who wrote the earliest known account of this. He
described a stone called \textit{Lyngourion} in Greek or \textit{Lyncurium} in
Latin, which had the property of attracting straws and pieces of wood. We now
know that this attraction is undoubtedly the result of electrostatic charges
generated by temperature fluctuations, most likely in mineral tourmaline.
Theophrastus and other writers of the following two millennia were far more
interested in the origin of the stone and its possible therapeutic properties
than its physical explanations. Theophrastus suspected that Lyngourion was
formed from the urine of a wild animal, which was later identified by Pliny the
Elder as lynx.

What is certain, however, is that accounts of this phenomenon reached Europe
in around 1700. The first printed description of the effect probably dates back
to
1707 and can be found in a work by Johann Georg Schmidt~\cite{Lang2005},
\cite{Katzir2006}.

\begin{quote}
The brilliant Dr. Damius (as Schmidt wrote) \ldots told me that in 1703, the
Dutch brought a gemstone called tourmaline, turmale, or trip, which not only
attracted ashes from warm or burning coals like a magnet attracts iron, but
also repelled them \ldots and I have no doubt that when heated, it would
attract other things besides ash.
\end{quote}

This unusual phenomenon has attracted significant interest in Europe. It has
been 
discussed at length in scientific circles and in various writings on natural
phenomena and curiosities. However, no connection with electricity was 
established at
that time. It was not until 1747 that Swedish naturalist Carl von Linné
appeared to have been the first to associate the attractive and repulsive
forces
of tourmaline with electricity. However, it remained a mere assumption:
Linnaeus
relied on unreliable oral reports and had never personally examined the
``electric stone''. He even went so far as to claim that the stone did not
exhibit this behavior ``either through heating or friction''. In
fact, the electrical nature of the phenomenon was only revealed nine years
later by the German astronomer, mathematician, physicist and natural
philosopher Franz Ulrich Theodosius Aepinus~\cite{Lang2005},
\cite{Katzir2006}.

At the end of 1756, German physician, mineralogist and geologist Johann Gottlob
Lehmann informed physicist Aepinus about a stone that attracted and repelled
small objects. It remains unclear whether Lehmann himself suspected that this
was because of to an electrical effect. Aepinus quickly recognized
that
this was an electrical phenomenon that differed from ``ordinary
electrical
phenomena,'' in which a body exhibits uniform electrical characteristics
across its entire surface. In his experiments, he immersed tourmaline in hot
water to heat it and analyzed the stone's electricity immediately after its
removal. They found that the two opposite ends of tourmaline had opposite
electrical properties at the same time. Based on his Franklinian theory of
electricity, he became convinced that the crystal had either an excess or
deficit of electricity at its opposite poles~\cite{Katzir2006},
\cite{Home1976}.

British electricity researchers Benjamin Wilson and John Canton supported
this view and conducted additional experiments with tourmaline. However, these
were not the only examples. As soon as Aepinus' discovery became known to
electrical scientists in Western Europe, it generated considerable interest.
However, the interpretation of positive and negative electricity consumption
is
not
universally accepted. In France, electricians instead interpreted the
phenomenon using the electricity system employed by French experimental
physicist Jean-Antoine Nollet, which was based on electric atmospheres and
varying inflows and outflows -- but apparently with less
success~\cite{Katzir2006}, \cite{Home1976}.

Aepinus believed that this novel electrical effect was due to heat, that is the
increased temperature of the crystal. In doing so, he failed to notice the drop
in the temperature of the samples during his investigations. However, he
noticed the importance of
the heating process and its uniformity: when heated unevenly,
the poles reversed compared to their ``natural'' state, which had been
created by homogeneous heating. It was not until three years later that
English physicist John Canton realized that the effect was related to a change
in temperature rather than to heat itself. Tourmaline, he noted, ``only
releases electrical fluid when the heat increases or decreases, and only
absorbs it in this way''. He also found that ``during heating, it
exhibits positive electricity on one side and negative electricity on the
other, which is also true when it is removed from boiling water and cools down;
however, the side that is positive when heated becomes negative when cooled,
and the side that is negative becomes positive''~\cite{Katzir2006},
\cite{Canton1759}.

In a subsequent experiment, he proved that positive and negative charges
were equally strong. Canton also confirmed an earlier assumption that the
characteristics of tourmaline are independent of its external form. Through
experiments, he showed that when tourmaline is split into three parts, both the
orientation of its polar axis and its properties remain unchanged. In 1760,
Canton discovered that other gemstones could also become electrically charged
owing
to temperature fluctuations. Thus, this phenomenon is not limited to
tourmaline. Shortly thereafter, Wilson conducted more detailed experimental
analyses of various minerals and concluded that their
electrical
characteristics, similar to those of tourmaline, are primarily determined by
their internal structure and not by their external form~\cite{Katzir2006},
\cite{Priestley1767}.

The study of electricity in crystals received significant new impulses through
research by the French mineralogist René-Just Haüy, who studied the crystal
structure and its characteristics in depth. Between 1785 and his death in 1822,
he conducted numerous studies in this field, including analyses of the
electrification of crystals resulting from temperature fluctuations. Inspired
by the concept of magnetism introduced by French physicist Charles Augustin de
Coulomb and the parallel presented by Aepinus between this and pyroelectricity,
he suggested that ``every molecule of heated tourmaline should be regarded
as a tiny electrical body, one end of which is positively charged and the other
negatively charged''~\cite{Hauy1792}.

Following the development of a comprehensive theory of crystal structure in
1801, Haüy applied this assumption to the crystallographic characteristics of
crystals. He associated the polar electrical units with the ``integrated
molecules''. Since every fragment of tourmaline retains the electrical
properties of the original whole, this hypothesis seemed ``extremely
plausible'' to him. In Haüy’s crystal theory, the integrated molecules
(molécules intégrants) form the fundamental building blocks of the crystal and
are 
shaped according to the crystal system. These molecules form a seamless,
continuous material, thereby creating a chain of positive and negative poles.
However, only the outer poles generate a noticeable external effect, whereas
the
inner poles neutralize one another~\cite{Hauy1801}, \cite{Blondel1997}.

Haüy also investigated the relationship between crystal structure and
electrical activity in various crystals and later observed that asymmetric
crystals -- those with an unequal number of faces at their ends (hemihedral) --
acquire an electrical charge when subjected to temperature changes, whereas
symmetric crystals do not~\cite{Katzir2006}. In 1840, the French mineralogist
and crystallographer Gabriel Delafosse, a former student of Haüy, emphasized
the relevance of symmetry to the physics of crystals. By relating the polarity
of crystal molecules to their structure, he limited his teacher's molecular
hypothesis: only asymmetric molecules can be electrically
polarized~\cite{Delafosse1843}.

Although the basic phenomenon of pyroelectricity has been known since Canton’s
time, numerous observations remained a mystery for a long time. Today we
understand that the observed electricity is based on a complex effect resulting
from uniform and non-uniform heating, as well as the conductivity of the
crystal and its surroundings. Although researchers have had theories regarding
the
influence of these parameters since the discovery of the phenomenon, a
comprehensive theoretical explanation (particularly with regard to the
different types of temperature changes) became possible following the
discovery of piezoelectricity~\cite{Katzir2006}, \cite{Forbes1834}.

In the 1760s, Aepinus and Wilson discussed the behavior of large
crystals when there were heated unevenly. Of greater relevance to understanding
the
phenomenon, however, was the question of whether the effect was caused solely
by temperature fluctuations and whether it persisted once the sample had
reached its final temperature. In 1828, the French physicist Antoine
César Becquerel, and in 1834 the Scottish physicist James David Forbes, felt it
necessary to reconfirm that the effect depends on temperature changes and not
on the absolute temperature. Furthermore, when quantifying the strength of
the electrical effect during the cooling process, it was observed that it
reached
its maximum before the sample reached its final temperature. Forbes therefore
contradicted the earlier statement by Scottish physicist David Brewster and
explained that tourmaline merely maintains its electrical state as long as its
temperature changes~\cite{Forbes1834}. Possibly owing to the numerous variables
associated with the temporal evolution of the charge, following these
experiments the researchers turned their attention to other aspects, such as
the occurrence of the phenomenon in various crystals and their connection to
their structure~\cite{Katzir2006}.

In 1824, David Brewster introduced the term ``pyroelectricity'' and
investigated its occurrence in numerous minerals using a device he
developed. In 1836, German mineralogist Gustav Rose studied the
orientation of these effects in tourmaline samples from various regions. He
later collaborated with German physicist Peter Rieß on the pyroelectric
properties of other minerals. Starting in 1839, German physicist Wilhelm
Gottlieb Hankel devoted his doctoral thesis to pyroelectricity and made it his
area of expertise. He supplemented earlier investigations into the distribution
of electricity on crystal surfaces through heating or cooling with new
experimental and, partly quantitative methods. Using these, he improved upon
existing results and analyzed the pyroelectric properties of various types of
crystals~\cite{Katzir2006}.

Aepinus emphasized the distinctive nature of pyroelectricity, the only
phenomenon that produces an electrified substance with two opposite poles, an
electric dipole, which cannot be broken down into two separate monopoles. The
polarity of tourmaline inspired him to draw a parallel between pyroelectric
matter and magnets, which always exhibit dipolar properties. He explained that
he was ``impressed by the astonishing similarity between this stone
(tourmaline) and the magnet'' and, inspired by this observation, had
``investigated the similarities between magnetic and electric forces with
greater
enthusiasm''~\cite{Connor1979}. This analogy also fascinated later
physicists, who were thinking about the cause of this phenomenon. Haüy
suspected
that the ``molécules intégrantes'' were polarizable, much like
Coulomb’s molecular magnets, so that a pyroelectric crystal consists of small
dipoles, just like a magnet. Brewster, who assumed that tourmaline retains its
electricity, also compared it with a magnet. Even Forbes, who believed that
tourmaline does not retain its polarity permanently like magnets, referred to
this analogy~\cite{Katzir2006}, \cite{Hauy1801}.

The new concepts of electricity in non-conductors developed by the English
experimental physicist Michael Faraday, which were based on magnetic analogies,
provided an alternative theoretical framework for interpreting dipolar
electricity in crystals. In an encyclopedia article on pyroelectricity
published in 1860, British physicist William Thomson, an active and young
(but already established) professor of natural philosophy in Glasgow, used
Faraday’s concept of electrical polarity to formulate his hypothesis regarding
the cause of the phenomenon~\cite{Katzir2006}.

\begin{quote}
The most likely explanation for the pyroelectric properties of dipolar crystals
(as Thomson wrote) is that these bodies naturally possess the same kind of
electrostatic polarization, which Faraday\ldots has clearly demonstrated to
occur temporarily in solid and liquid non-conductors, and that they possess
this property to varying degrees at different temperatures. The inductive
effect that this electropolar state of the substance has on the matter
surrounding the body induces a superficial electrification that perfectly
balances its electrical force at all points of the external matter\ldots When
the temperature of the substance changes, its electropolarisation changes
simultaneously, while the masking superficial electrification follows the
charge only slowly -- more or less slowly -- depending on the greater or lesser
resistance that the substance or its surface offers to electrical
conduction~\cite{Thomson1878}.
\end{quote}

Thomson’s proposal builds on Haüy’s hypothesis and can be regarded as a further
development of it, although it differs in two key respects. First, Thomson
made no assumptions about the origin or exact location of the polarity within
the crystal. Second, and crucially, he assumes for the first time that
polarity persists permanently, even when the crystal and its parts appear
electrically neutral. The parallel relationship between polar electricity and
polar
magnetism is already implicit in the theory of electromagnetism. As the
mathematical relationships originally developed for magnetism were also
applicable to electrical polarity, this analogy gained significance. When
Thomson reintroduced his hypothesis in 1878, he used this analogy to predict
the existence of a reverse pyroelectric (electrocaloric) effect, namely a
change in temperature caused by electrification~\cite{Katzir2006}.

In summary, it can be said that around 1880, when the Curie brothers (both
French physicists) were considering the source of pyroelectricity and
piezoelectricity, it was known that pyroelectricity is linked to a change in
temperature in hemihedral, that is, asymmetrical, crystals. It is known that
the
generation and distribution of electricity are closely related to the
crystal structure. Haüy conducted numerous experiments to investigate
these relationships for various crystal types. Hankel, a specialist
in
such
experimental investigations, was still active at the time. The phenomenon
lacked theoretical explanation or justification, but Gaugain and Thomson made
initial progress in this direction. Gaugain developed quantitative laws
governing the generation of electric charges in pyroelectric crystals, which
showed that the strength of the effect depends on the temperature difference,
specific properties of the crystal type, and its surface. Following Haüy et.
\,al.,
Thomson proposed explaining the effect of the internal polarity
of the crystal~\cite{Katzir2006}.

In the second half of the 19th century and in the first decades of the
20th century, seven Nobel laureates -- Wilhelm Conrad
Röntgen, Pierre Curie, Gabriel Lippmann, Heike Kammerlingh Onnes, Erwin
Schrödinger, Archer J. P. Martin, 
and Max Born -- published papers on pyroelectricity, 
underlining the
scientific significance of this field during that era.

In 1920, the American physicist Joseph Valasek investigated the properties of
sodium tartrate, also known as Rochelle salt, and in doing so
discovered ferroelectricity. With the discovery of ferroelectricity, interest
in pyroelectricity almost completely disappeared, until Yeou Ta published a
paper in 1938 that marked the beginning of the major boom in
this field to this day. Ta, a chemist at Sorbonne in Paris, proposed the use of
tourmaline crystals as IR sensors in spectroscopy. During and immediately after
the Second World War, some studies on pyroelectric IR detectors were carried
out in the UK, USA, and Germany, although the results were published only in
classified documents~\cite{Lang2005}.

In 1962, Cooper performed the first detailed analysis of the behavior
of fast IR detectors and conducted experiments using barium titanate. In the
same year, he proposed the use of pyroelectric devices to measure temperature
changes in the range of just $\SI{0,2}{\micro\kelvin}$. This marked the
beginning of an explosive increase in theoretical studies, fundamental
measurements and applications: since 1960, more than 8,500 articles on
pyroelectricity have been published~\cite{Lang2005}.

The historical
development
described in Section~\ref{sec:Entdeckung} provides
a physical description of pyroelectricity within the crystal
physics framework.
In Section~\ref{sec:Physik}, the quantities of the
spontaneous dipole moment
of the unit cell, spontaneous polarization and pyroelectric coefficient
are introduced and formally described within the framework
of crystal physics.

\section{The physical principle behind pyroelectricity}
\label{sec:Physik}

\begin{figure}[h]
\centering
\includegraphics[width=\linewidth]
{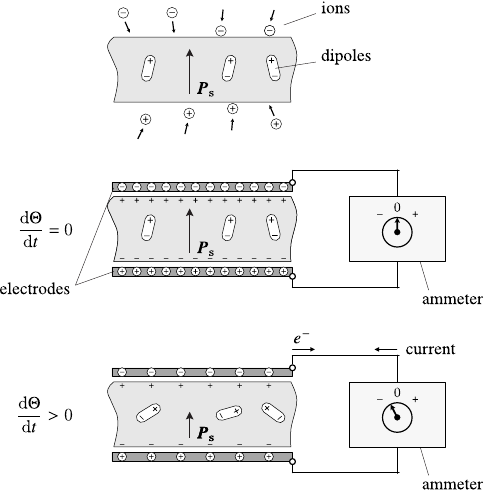}
\caption{When a pyroelectric crystal with an intrinsic dipole moment (top) is
incorporated into a circuit with electrodes attached to each surface (middle),
an increase in temperature $\Theta$ causes the spontaneous polarisation
$\boldsymbol{P}_{\mathrm{s}}$ to decrease, as the dipole moments become, on
average, smaller in magnitude. The horizontal inclination of the dipoles, shown
below, illustrates this effect. A current flows to compensate for the change in
bound charge that accumulates at the crystal edges (adapted
from~\cite{Lang2005}).}
\label{fig:Pyroelektrizitaet}
\end{figure}

One of the least well-known properties of solids, pyroelectricity, is defined
as the
temperature dependence of spontaneous polarization in certain
anisotropic solids~\cite{Lang2005}.
To understand the significance of this definition and the nature of the
pyroelectric effect, consider a simple example of a thin, parallel-sided
sample of a material, such as a tourmaline crystal or a barium titanate
ceramic disc, cut such that its crystallographic axes of symmetry are
perpendicular to the flat surfaces (see Figure~\ref{fig:Pyroelektrizitaet}
above).

The spontaneous dipole moment describes the presence of a dipole moment in the
absence of an external electric field below a certain transition temperature,
known as the Curie temperature~\cite{Ballas2025a}. The vector sum of all $j$
electronic and atomic dipole moments $\boldsymbol{p}_j$ of a crystal’s unit
cell yields its permanent (spontaneous) dipole moment
$\boldsymbol{p}_{\mathrm{s}}$. The following applies:
\begin{equation}
	\boldsymbol{p}_{\mathrm{s}}=
	\sum_{j}^{N}\boldsymbol{p}_j
	\label{eq:spontanes Dipolmoment}
\end{equation}
The upper summation index $N$ denotes the number of the dipole
moments. The spontaneous dipole moment $\boldsymbol{p}_{\mathrm{s}}$ of the
unit cell represents the microscopic quantity of the crystal~\cite{Sonin1974}.

The dipole moment per unit volume of the material is referred to as the
spontaneous polarization $\boldsymbol{P}_{\mathrm{s}}$. 
$\boldsymbol{P}_{\mathrm{s}}$ is always non-zero in a pyroelectric material,
exists in the absence of an applied electric field, and corresponds to a layer
of bound charge $Q$ on each flat surface of the sample. Nearby free charges
such as electrons or ions are attracted by the sample (see
Figure~\ref{fig:Pyroelektrizitaet})~\cite{Lang2005}.

Although the spontaneous dipole moment is a microscopic quantity, a
macroscopic crystal (dielectric) with volume $V$, whose opposite faces,
separated by a distance $\boldsymbol{l}$, possess a layer of bound charge $Q$,
it can be described as follows:
\begin{equation}
	\boldsymbol{P}_{\mathrm{s}}=
	\frac{Q\boldsymbol{l}}{V}
	\label{eq:spontane}
\end{equation}
(see also Equation~(\ref{eq:sP2}))~\cite{Zheludev1971a}). In most crystals, the
direction of the spontaneous polarisation $\boldsymbol{P}_{\mathrm{s}}$
coincides with the spontaneous dipole moment $\boldsymbol{p}_{\mathrm{s}}$ of
the unit cell. The SI unit of polarisation is
$\SI{1}{\ampere\second.\metre.\metre^{-3}}=\SI{1}{\coulomb.\metre^{-2}}$~\cite
{Sonin1974}.

\begin{figure}
\centering
\includegraphics[width=\linewidth]
{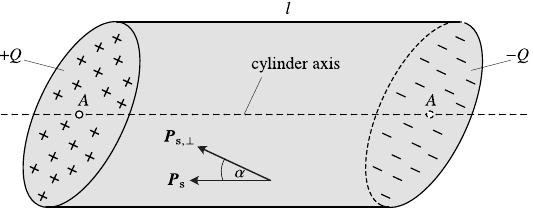}
\caption{An inclined circular cylinder (cylindrical length $l$, end faces $A$)
illustrating the relationship between the spontaneous polarisation
$\boldsymbol{P}_{\mathrm{s}}$ and the surface charge
$\sigma$ (adapted from~\cite{Sonin1974}).}
\label{fig:Messung Polarisation}
\end{figure}

Figure~\ref{fig:Messung Polarisation} illustrates how
polarization can be measured. The spontaneous polarization vector
$\boldsymbol{P}_{\mathrm{s}}$ was assumed to be oriented parallel to the
cylinder axis. The electric centers of mass of the positive and negative
charges in the crystal did not coincide. Therefore, one can visualize the
system
of dipoles as being replaced by charges $+Q$ and $-Q$ on the end faces of
the sample. For the normal component of spontaneous polarization, the
following equation applies~\cite{Sonin1974}:
\begin{equation*}
	P_{\mathrm{s},\perp}=
	P_{\mathrm{s}}\cos\alpha
\end{equation*}
The quantity $\alpha$ represents the angle between $P_{\mathrm{s}}$ and the
normal component $P_{\mathrm{s},\perp}$. According to
Definition~\ref{eq:spontane}, the following applies in relation to volume
$V$ of the sample
\begin{equation}
	P_{\mathrm{s}}=
	\frac{Ql}{V}\text{.}
	\label{eq:sP2}
\end{equation}
The normal component is therefore given by:
\begin{equation*}
	P_{\mathrm{s},\perp}=
	\frac{Ql}{V}\cos\alpha
\end{equation*}
The charge density $\sigma$ on the end faces of the circular cylinder is given
by
\begin{equation*}
	\sigma=\frac{Q}{A}=
	\frac{Ql}{V}\cos\alpha\text{,}
\end{equation*}
from which it immediately follows
\begin{equation*}
	\sigma=
	P_{\mathrm{s},\perp}\text{,}
\end{equation*}
the normal component of $P_{\mathrm{s}}$ is equal to the charge density on
the crystal surface~\cite{Sonin1974}.

Despite spontaneous polarization, the surface charge of polar crystals is
normally zero, as it is compensated for by free charges within the crystal or
in the surrounding environment. It can only be measured directly on the
surface of a fresh fracture, before compensation occurs via conductivity or
contamination~\cite{Sonin1974}.

Pyroelectric crystals exhibit spontaneous polarization
$\boldsymbol{P}_{\mathrm{s}}$, the values of which can be varied by uniform
heating. Such a change in spontaneous polarization is responsible for
pyroelectric phenomena~\cite{Zheludev1971a}. 
For small temperature changes $\Delta\Theta$, the magnitude of the pyroelectric
effect is proportional to the change in the polarization
$\Delta\boldsymbol{P}_{\mathrm{s}}$. The proportionality factor
$\boldsymbol{p}$ denotes the pyroelectric coefficient. The following holds:
\begin{equation}
	\Delta\boldsymbol{P}_{\mathrm{s}}=
	\boldsymbol{p}\,\Delta\Theta
	\label{eq:koeffizient}
\end{equation}
The change in polarisation is due to two distinct effects~\cite{Zheludev1971a}:

\begin{enumerate}
\item the primary or true pyroelectric effect
\item the secondary or false pyroelectric effect
\end{enumerate}

The primary pyroelectric effect describes the change in the polarization
$\Delta\boldsymbol{P}_{\mathrm{s}}$, which is not related to the piezoelectric
effect. Every crystal tends to change its dimensions when heated, which is
inevitably linked to the piezoelectric effect. In a free
crystal, this effect is due to the resulting deformation; in a clamped crystal,
it is due to thermal stress~\cite{Zheludev1971a}.

The change in the polarization $\Delta\boldsymbol{P}_{\mathrm{s}}$ associated
with
the piezoelectric effect characterizes the secondary pyroelectric effect.
Therefore, the
total effect is the sum of the primary and secondary pyroelectric
effects~\cite{Zheludev1971a}.

To calculate the secondary pyroelectric effect, we consider the deformation of
a free (unclamped) crystal when it is heated uniformly from temperature
$\Theta_{1}$ to temperature $\Theta_{2}$. According to
Equation~(\ref{eq:spontane}),
the following equation is applied:
\begin{align}
	\boldsymbol{P}^{\prime\prime}_{s}(\Theta_1)&=\frac{Q}{V}\,\boldsymbol{l}_
{\Theta_{1}} \notag \\
\boldsymbol{P}^{\prime\prime}_{s}(\Theta_2)&=\frac{Q}{V}\,\boldsymbol{l}_
{\Theta_{2}} \notag
\end{align}
The superscript notation $^{\prime\prime}$ is intended to emphasize that this
refers to the secondary pyroelectric effect. The change in the polarization
$\Delta\boldsymbol{P}^{\prime\prime}_{\mathrm{s}}$ is given by
Equation~(\ref{eq:koeffizient}) as
\begin{equation*}
	\Delta\boldsymbol{P}^{\prime\prime}_{\mathrm{s}}=
	\frac{Q}{V}\Delta\boldsymbol{l}=
	\boldsymbol{p}^
{\prime\prime}\Delta\Theta\quad\text{with}\quad\Delta\Theta=\Theta_2-\Theta_1
\end{equation*}
The quantity $\boldsymbol{p}^{\prime\prime}$ denotes the secondary pyroelectric
coefficient (Table~\ref{tab:Pyrokoeffizienten}).

In practice, it is difficult to distinguish between the primary and secondary
pyroelectric effects, because experimental measurements reflect the overall
effect.
\begin{equation*}
	\Delta\boldsymbol{P}_{\mathrm{s}}=
	(\boldsymbol{p}^{\prime}+\boldsymbol{p}^{\prime\prime})\,\Delta\Theta
\end{equation*}
The quantity $\boldsymbol{p}^{\prime}$ denotes the primary pyroelectric
coefficient and $\boldsymbol{p}^{\prime\prime}$ denotes the
secondary pyroelectric coefficient (see also
Table~\ref{tab:Pyrokoeffizienten})~\cite{Zheludev1971a}.

\begin{table*}
\renewcommand{\arraystretch}{1.2}
\newcolumntype{P}[1]{>{\centering\arraybackslash}p{#1}}
\caption
{Primary, secondary and total pyroelectric coefficients of various materials.
(Unit: \SI{1}{\micro\coulomb.\metre^{-2}.\kelvin^{-1}})  (according
to~\cite{Lang2005})}
\label{tab:Pyrokoeffizienten}
\centering
\begin{tabular}{P{3.5cm}P{3.8cm}P{2.16cm}P{2.16cm}P{2.16cm}}
\hline\noalign{\smallskip}
\textbf{Material}
&    \textbf{Name}
&    \textbf{Primary coefficient}
&    \textbf{Secondary coefficient}
&    \textbf{Total coefficient}
	\\		
	\noalign{\smallskip}\hline\noalign{\smallskip}	
	%
	\textbf{Ferroelectrics}         
&           
&              
&                      
&					
	\\
	\textbf{\textit{Ceramics}    }
&          
&            
&                        
&							
	\\
	BaTiO\textsubscript{3}    
&   barium titanate 	 
&	\num{-260} 			
&	+60 					
& 	\num{-200} 		
	\\
	PbZr\textsubscript{0.95}Ti\textsubscript{0.05}O\textsubscript{3}
&   lead zirconate titanate 	 	
&	\num{-305,7} 		
&	+37.7 					
&	\num{-268} 		
	\\
	\textbf{\textit{Crystal}}
&
&
&
& 
	\\
	LiNbO\textsubscript{3}	
&	lithium niobate 	 
&	\num{-95,8} 			
&	+\num{12,8} 					
& 	\num{-83} 		
	\\
	LiTaO\textsubscript{3}	
&	lithium tantalate 	 
&	\num{-175} 			
&	\num{-1} 					
& 	\num{-176} 		
	\\
	Pb\textsubscript{5}Ge\textsubscript{3}O\textsubscript{11}	
&	lead nitrate 	 
&	\num{-110,5} 			
&	+\num{15,5} 					
& 	\num{-95} 		
	\\
Ba\textsubscript{2}NaNb\textsubscript{5}O\textsubscript{15}
&	barium sodium niobate 
&	\num{-141,7} 			
&	+\num{41,7} 					
& 	\num{-100}
	\\
Sr\textsubscript{0.5}Ba\textsubscript{0.5}Nb\textsubscript{2}O\textsubscript
{6}	
&	strontium barium niobate 	 
&	\num{-502} 			
&	\num{-48} 					
& 	\num{-550}
	\\
(CH\textsubscript{2}CF\textsubscript{2})\textsubscript{n}	
&	polyvinylidene fluoride 	 
&	\num{-14} 			
&	\num{-13} 					
& 	\num{-27}
	\\
C\textsubscript{6}H\textsubscript{17}N\textsubscript{3}O\textsubscript{10}S
&	triglycine sulphate 	 
&	+\num{60} 			
&	\num{-330} 					
& 	\num{-270}
	\\
	\noalign{\smallskip}\hline\noalign{\smallskip}
\textbf{None-ferroelectrics}
&
&
&
& 
	\\
\textbf{\textit{Crystal}}
&
&
&
& 
	\\
CdSe
&	cadmium selenide 	 
&	\num{-2,94} 			
&	\num{-0,56} 					
& 	\num{-3,5}
	\\
CdS
&	cadmium sulphide 	 
&	\num{-3,0} 			
&	\num{-1,0} 					
& 	\num{-4,0}
	\\
ZnO
&	Zinc oxide 	 
&	\num{-6,9} 			
&	\num{-2,5} 					
& 	\num{-9,4}
	\\
Tourmaline
&	 	 
&	\num{-0,48} 			
&	\num{-3,52} 					
& 	\num{-4,0}
	\\
Li\textsubscript{2}SO\textsubscript{4} $\cdot$ 2H\textsubscript{2}O
&	 lithium sulphate	 
&	+\num{60,2} 			
&	+\num{26,1} 					
& 	+\num{86,3}
	\\
	\noalign{\smallskip}\hline\noalign{\smallskip}
\end{tabular}
\end{table*}

The vector equation
\begin{equation*}
\Delta\boldsymbol{P}_{\mathrm{s}}=
(\boldsymbol{p}^{\prime}+\boldsymbol{p}^{\prime\prime})\,\Delta\Theta
\end{equation*}
can be expressed in terms of the components of the vectors:
\begin{equation*}
	\Delta P_{\mathrm{s}i}=
	(p^{\prime}_i+p^{\prime\prime}_i)\,\Delta\Theta
	\quad\text{mit}\quad i=1,2,3
\end{equation*}
In differential form, this gives
\begin{equation*}
p_i=\dfrac{\partial P_{\mathrm{s}i}}{\partial\Theta}\text{,}
\end{equation*}
where $p_i$ is the total pyroelectric coefficient
($p_i=p^{\prime}_i+p^{\prime\prime}_i$)~\cite{Zheludev1971a}.

Ferroelectric materials are a subclass of
pyroelectric
materials because they possess spontaneous polarization that is fundamentally
temperature-dependent~\cite{Zheludev1971a}, \cite{Lines1977}. The
key difference is 
that the direction of this polarization can be reversed by an external electric
field~\cite{Lines1977}. In pyroelectric non-ferroelectric materials
such, as tourmaline or (cadmium sulfide) CdS , such a reversal is not
possible, as the
required
coercive field exceeds the breakdown field strength~\cite{Lines1977}.
Ferroelectrics typically
exhibit high pyroelectric coefficients (see
Table~\ref{tab:Pyrokoeffizienten}) and are therefore particularly important for
infrared sensors and
pyroelectric energy converters~\cite{Lang2005}. Table
\ref{tab:Pyrokoeffizienten} provides an
overview of the common
pyroelectric materials.

Table~\ref{tab:Pyrokoeffizienten}
lists several
key
relationships. In ferroelectrics
(BaTiO\textsubscript{3}, PZT), the primary
pyroelectric coefficient typically dominates,
whereas in non-ferroelectrics such as tourmaline, the secondary component
predominates.
A negative coefficient indicates a decrease in spontaneous polarization with
increasing temperature, while a positive coefficient (e.g. lithium sufhate)
indicates an increase. This is particularly pronounced in (triglycine
sulfate) TGS.
Although the primary coefficient is positive
($\SI{60}{\micro\coulomb.\metre^{-2}.\kelvin^{-1}}$),
the
strongly
negative secondary component
($\SI{-330}{\micro\coulomb.\metre^{-2}.\kelvin^{-1}}$)
results in a negative
overall coefficient. Despite its low coefficient
($\SI{-4,0}{\micro\coulomb.\metre^{-2}.\kelvin^{-1}}$), tourmaline has
historical
significance
and continues to be used as a reference material for
spectroscopy~\cite{Lang2005}.

There are a wide variety of pyroelectric materials, including minerals such as
tourmaline, single crystals such as triglycine sulfate, ceramics such as lead
zirconate titanate, polymers such as polyvinylidene fluoride, and even
biological materials such as collagen~\cite{Lang2005}. 
Table~\ref{tab:Pyrokoeffizienten} provides an overview of the common
pyroelectric
materials.

\section{Conclusion}

The history of pyroelectricity illustrates the
long journey from ancient observations of nature to precisely
describing the physical properties of modern functional materials. While the
attractive effect of heated tourmaline
remained a mystery for over two millennia, it was only a systematic research in
the 18th and 19th centuries, beginning with Aepinus and Canton and
culminating
in the theoretical work of William Thomson, which enabled it to be classified
as an electrical phenomenon.

Physical analysis showed that pyroelectricity is inextricably related to the
internal structure of anisotropic solids. The fundamental quantity here is the
spontaneous polarization $\boldsymbol{P}_{\mathrm{s}}$, which results from the
vector sum of the microscopic dipole moments $\boldsymbol{p}_{\mathrm{s}}$. 
A key finding of the theoretical analysis is that the
measurable total effect consists of the sum of two components: $\Delta
P_{\mathrm{s}}=\left(p^{\prime}+p^{\prime\prime}\right)\Delta\Theta$ Here,
where the
primary pyroelectric coefficient $p^{\prime}$ describes the intrinsic change in
polarization at constant volume, while the secondary coefficient
$p^{\prime\prime}$ represents the coupling via thermal expansion and the
associated piezoelectric effect. As the material overview
(Table~\ref{tab:Pyrokoeffizienten}) illustrates, these components can exhibit
different signs and magnitudes depending on the crystal structure.

Although interest in the field waned temporarily following the discovery of
ferroelectricity in 1920, pyroelectricity has experienced unprecedented
renaissance since the 1960s. With over 8,500 publications and widespread
application in infrared sensor technology, spectroscopy and non-contact
temperature measurement (down to the range of $\SI{0,2}{\micro\kelvin}$), it
is now an indispensable component of modern instrumentation.

Future developments are expected to take three main directions. First, the
integration of pyroelectric thin films into
micro-electromechanical systems (MEMS) is gaining importance, as it enables
miniaturized
IR detectors without external cooling~\cite{Starman2025}, \cite{Ng2020}.
Second, 
the
discovery
of pyroelectric properties in biological materials, particularly collagen
and bone, opens up new prospects for biomedical
sensors~\cite{Wang2026}, \cite{Yuan2024}. Third, pyroelectric
energy
harvesting
is in the spotlight for
the conversion of temperature fluctuations into electrical energy for
environmental sensing and IoT applications~\cite{Mondal2023},
\cite{Mohammadnia2023}.
Fundamental
open
questions
concern first-principles modelling of the primary coefficient as well as
the characterization of biological pyroelectric materials. Thus, even more than
2,000 years after
its
first
mention by Theophrastus, pyroelectricity remains an active field of research in
materials science with growing technical relevance.

\backmatter

\bibliography{Pyroelectricity}

\begin{thebibliography}{22}
\bibitem[\protect\citeauthoryear{Canton}{1759}]{Canton1759}
\begin{barticle}
\bauthor{\bsnm{Canton}, \binits{J.}}:
\batitle{{{XXXVIII}. An attempt to account for the regular diurnal variation of
  the horizontal magnetic needle; and also for its irregular variation at the
  time of an Aurora Borealis}}.
\bjtitle{Philosophical Transactions of the Royal Society of London}
\bvolume{51},
\bfpage{398}--\blpage{445}
(\byear{1759})
\end{barticle}
\endbibitem
\end{thebibliography}

\end{document}